\documentclass[aps,prd,onecolumn,nofootinbib,groupedaddress,superscriptaddress,11pt]{revtex4}  
\usepackage{graphicx}
\usepackage{epstopdf}
\usepackage{amsmath}
\usepackage{amsfonts}
\usepackage{amssymb}
\usepackage{appendix}
\usepackage{comment}
\usepackage{bbold}
\usepackage{color}
\usepackage{slashed}
\usepackage{subfigure}
\usepackage{setspace}
\usepackage{footnote}
\usepackage{multirow}
\usepackage{longtable}
\usepackage{braket}
\usepackage{appendix}
\usepackage{verbatim}
\usepackage{xcolor}
\newcommand{\beq}{\begin{equation}}
\newcommand{\eeq}{\end{equation}}

\usepackage{hyperref}
\definecolor{nicered}{rgb}{0.7,0.1,0.1}
\definecolor{nicegreen}{RGB}{53,170,102}
\definecolor{niceblue}{rgb}{0.117,0.5625,1.0}
\definecolor{nicepurple}{RGB}{127, 38, 222}
\hypersetup{colorlinks,
			citecolor      = nicepurple, 
			linkcolor       = niceblue, 
			urlcolor        = nicegreen, 
			anchorcolor = niceblue}

\begin{document}

\singlespacing

{\hfill FERMILAB-PUB-22-663-T}

\title{Majorana versus Dirac Constraints on the Neutrino Dipole Moments}

\author{Andr\'{e} de Gouv\^{e}a} 
\affiliation{Northwestern University, Department of Physics \& Astronomy, 2145 Sheridan Road, Evanston, IL 60208, USA}
\author{Giancarlo Jusino S\'anchez} 
\affiliation{Northwestern University, Department of Physics \& Astronomy, 2145 Sheridan Road, Evanston, IL 60208, USA}
\author{Pedro A.N. Machado}
\affiliation{Particle Theory Department, Fermilab, P.O. Box 500, Batavia, IL 60510, USA}
\author{Zahra Tabrizi} 
\affiliation{Northwestern University, Department of Physics \& Astronomy, 2145 Sheridan Road, Evanston, IL 60208, USA}

\begin{abstract}
%%%%%%%%%%%%%%
Massive neutrinos are guaranteed to have nonzero electromagnetic moments and, since there are at least three neutrino species, these dipole moments define a matrix. Here, we estimate the current upper bounds on all independent neutrino electromagnetic moments, concentrating on Earth-bound experiments and measurements with solar neutrinos, including the very recent results reported by XENONnT. We make no simplifying assumptions and compare the hypotheses that neutrinos are Majorana fermions or Dirac fermions. 
In particular, we fully explore constraints in the Dirac-neutrino parameter space. 
Majorana and Dirac neutrinos are different; for example, the upper bounds on the magnitudes of the elements of the dipole moment matrix are weaker for Dirac neutrinos, relative to Majorana neutrinos. 
The potential physics reach of next-generation experiments also depends on the nature of the neutrino. We find that a next-generation experiment two orders of magnitude more sensitive to the neutrino electromagnetic moments via $\nu_{\mu}$ elastic scattering may discover that the neutrino electromagnetic moments are nonzero if the neutrinos are Dirac fermions. Instead, if the neutrinos are Majorana fermions, such a discovery is ruled out by existing solar neutrino data, unless there are more than three light neutrinos.  

%%%%%%%%%%%%%%%%%
\end{abstract}

%\pacs{13.35.Hb,14.60.St,11.30.Fs}

\maketitle

%%%%%%%%%%%%%%%%%%
\section{Introduction}
\label{sec:Intro}
\setcounter{equation}{0}
%%%%%%%%%%%%%%%%%%

While the neutrino charge is zero, massive neutrinos are guaranteed to have a nonzero electromagnetic dipole moment. In the absence of new interactions, the neutrino magnetic dipole moment is generated at the one loop level and is of order $10^{-20}(m_{\nu}/0.1~\rm eV) \mu_B$ \cite{Marciano:1977wx,Lee:1977tib,Fujikawa:1980yx}, where $\mu_B$ is the Bohr magneton. This is several orders of magnitude beyond the sensitivity of current and near future experimental probes. The measurement of a nonzero neutrino electromagnetic dipole moment would imply more new physics in the neutrino sector. 

The nature of the neutrino dipole moment depends on whether neutrinos are Majorana fermions or Dirac fermions. It is well known that diagonal dipole moments for Majorana fermions are forbidden and hence these only have transition dipole moments. Dirac fermions, instead, are allowed to have both diagonal and transition dipole moments. We review this carefully in Section~\ref{sec:theory}, concentrating on the differences between Dirac and Majorana neutrinos. We also discuss how Majorana neutrinos can ``mimic'' Dirac neutrinos in the presence of new light neutral fermions.

Nonzero neutrino electromagnetic dipole moments contribute to neutrino--matter scattering, as we discuss in more detail in Section~\ref{sec:observables}. Precision measurements of neutrino scattering, therefore, allow one to constrain their magnitudes. Experiments with reactor antineutrinos and solar neutrinos, for example, exclude effective dipole moments larger than a few times $10^{-11}\mu_B$~\cite{Beda:2012zz,Borexino:2017fbd}.
 In Section~\ref{sec:experiments} we list the current laboratory constraints. In the near and intermediate future, better laboratory sensitivity is expected~(see, for example, \cite{Billard:2018jnl,Miranda:2019wdy,Baxter:2019mcx,Mathur:2021trm,Aalbers:2022dzr}). There are also indirect constraints on the neutrino electromagnetic dipole moments from astrophysical processes \cite{Heger:2008er,Diaz:2019kim,Lattimer:1988mf,Raffelt:1999gv}. We comment on those briefly in Section~\ref{sec:experiments}. Here, we concentrate on laboratory constraints, which we view as complementary to the indirect astrophysical bounds.

Since there are at least three different neutrino flavors, a more careful examination of the experimental data is required. Different experiments constrain different combinations of the neutrino dipole moments. This implies that (a) some combinations of dipole moments are less constrained and (b) one can obtain qualitatively different bounds on the neutrino dipole moments by combining information from different experiments. The interplay of the different data sets also depends on whether neutrinos are Dirac or Majorana fermions. Here we estimate the current bounds on all neutrino dipole moments, taking all possible correlations into account, for both Dirac neutrinos and Majorana neutrinos. We also discuss expectations for future experimental searches. We find, in particular, that expectations depend strongly on whether neutrinos are Majorana or Dirac fermions.  These results are presented and discussed in Sec~\ref{sec:results}.

The constraints reported by the experiments are in the form of upper limits on the magnitude of some effective magnetic moment $|\mu^{\rm eff}|$ (see Section~\ref{sec:observables}). These constraints can be translated into the fundamental electromagnetic dipole moments. In recent years, there have been many efforts connecting these constraints to the parameters of the Lagrangian~(see, for example, \cite{Beacom:1999wx,Grimus:2000tq,Joshipura:2002bp,Grimus:2002vb,Canas:2015yoa,Billard:2018jnl,Miranda:2019wdy,Miranda:2021kre,AristizabalSierra:2021fuc,Cadeddu:2020lky,AtzoriCorona:2022qrf,Khan:2022bel,A:2022acy}). In most of these studies, special attention was dedicated to the Majorana-neutrino hypothesis. In this case, relative to the Dirac-neutrino hypothesis, there are fewer parameters and the analysis is computationally simpler. For the Dirac-neutrino hypothesis, it is often the case that only constraints on the diagonal magnetic moments are considered in the literature. Here, we present the results of a comprehensive analysis, treating all the parameters as independent from one another. We also discuss is some detail what information is, in principle, experimentally accessible. We make use the experimental data of current solar, reactor and accelerator experiments, including the most recent results from XENONnT \cite{XENON:2022mpc} (also discussed, very recently, in \cite{Khan:2022bel,A:2022acy}), and speculate on the impact of a future accelerator experiment capable of constraining the neutrino dipole moment using a $\nu_{\mu}$ ``beam.'' We find that such a future experiment has the potential to make a discovery even when its sensitivity is significantly weaker than the current solar constraints. However, this statement is only true, assuming there are no new light particles, if the neutrinos are Dirac fermions. 

The fact that electromagnetic dipole moments and masses are correlated -- both require chirality violation -- also allows one to estimate how large the neutrino dipole moments could be. In a nutshell, generic new physics that induces nonzero neutrino dipole moments will also contribute to the neutrino masses. If one assumes the new-physics contribution to the neutrino masses is not much larger than the known values, one can place mostly model-independent bounds on the neutrino dipole moments \cite{Bell:2005kz,Bell:2006wi}.   In \cite{Bell:2005kz}, Bell and collaborators argued that, modulo fine-tuning among different contributions to the neutrino masses, neutrino dipole moments are guaranteed to be less than, roughly, $10^{-15}(m_{\nu}/0.1~\rm eV)\mu_B$ if the neutrinos are Dirac fermions. The equivalent upper bound on Majorana neutrinos is a lot weaker. For example, if there is new physics at the weak scale, it is possible to identify scenarios that saturate the current experimental constraints (see, e.g., \cite{Bell:2006wi} and references therein and \cite{Lindner:2017uvt} for a more recent discussion). We return to these issues in Section~\ref{sec:conclusions}, where we also summarize our results and offer other concluding remarks.

%This manuscript is organized as follows. In Section~\ref{sec:theory} we review the formalism that describe neutrino magnetic moments, concentrating on the differences between Dirac and Majorana neutrinos. We also discuss how much Majorana neutrinos can ``mimic'' Dirac neutrinos in the presence of new light fermions. In Section~\ref{sec:experiments} we briefly review how the neutrino magnetic moment is manifests itself in neutrino scattering experiments and list the current constraints. There we also discuss our strategy for using existing constraints, and combinations thereof, in order to place bounds on Dirac and Majorana neutrino magnetic moments. Our results are presented and discussed in Sec~\ref{sec:results}. Section~\ref{sec:conclusions} contains some final remarks and observations. 

%%%%%%%%%%%%%%%%%%
\section{The Electronmagnetic Dipole Moment Matrix}
\label{sec:theory}
\setcounter{equation}{0}
%%%%%%%%%%%%%%%%%%

Given two left-handed Weyl fermions $\chi_a$ and $\chi_b$ with zero electric charge, one can write down the following gauge and Lorentz invariant dimension-five operator that couples the fermions to the electromagnetic field strength $F^{\mu\nu}$: 
\begin{equation}
\mathcal{O} = \frac{1}{\Lambda} (\chi_a)^{\beta}
\left[(\sigma_{\mu})_{\beta\dot\alpha}(\bar{\sigma}_{\nu})^{\dot\alpha\alpha}-(\sigma_{\nu})_{\beta\dot\alpha}(\bar{\sigma}_{\mu})^{\dot\alpha\alpha}\right]
(\chi_b)_{\alpha}F^{\mu\nu},
\label{eq:weyl}
\end{equation}
making use of the standard $\alpha,\dot\alpha=1,2$ notation for Weyl fermions, along with the $\epsilon^{\alpha\beta}=-\epsilon^{\beta\alpha}$ metric for raising and lowering spinor indices (there is an equivalent metric for dotted indices), the four-vector $\sigma_{\mu}, \bar{\sigma}_{\mu}$ $2\times 2$--matrices, while $\Lambda$ denotes an arbitrary energy scale. 
It is easy to show that, now omitting spinor indices, $\chi_a\sigma_{\mu}\bar{\sigma}_{\nu}\chi_b=\chi_b\sigma_{\nu}\bar{\sigma}_{\mu}\chi_a$ so Eq.~(\ref{eq:weyl}) is antisymmetric upon the exchange $a\leftrightarrow b$. This means that Eq.~(\ref{eq:weyl}) for $a=b$ vanishes exactly.

If neutrinos are Majorana fermions, each neutrino mass eigenstate $\nu_i$ (with mass $m_i$, $i=1,2,\ldots, N$, and $N$ is the number of neutrinos) can be represented as a two-component left-handed Weyl fermion and the following Lagrangian describes the neutrino--photon interactions at dimension five:
\begin{equation}\label{eq:LagM}
{\cal L}_{\rm M} = \frac{1}{2}\mu_{ij} \nu_i\sigma_{\mu\nu}\nu_jF^{\mu\nu} + H.c.,
\end{equation}
where $4\sigma_{\mu\nu}\equiv\sigma_{\mu}\bar{\sigma}_{\nu}-\sigma_{\nu}\bar{\sigma}_{\mu}$. Here, $\mu_{ij}=-\mu_{ji}$ are complex constants that define the neutrino electromagnetic dipole moment matrix. There are $(N^2-N)/2$ complex, independent $\mu_{\ij}$. 
In the case of three neutrinos, the dipole moment matrix is parameterized by 6 real parameters: $\mu_{ij}=|\mu_{ij}|e^{i\xi_{ij}}$, $ij=12,13,23$.

If neutrinos are Dirac fermions, each neutrino mass eigenstate can be represented as a pair of two-component left-handed Weyl fermions, $\nu_i$, and $\nu_i^c$. 
In our notation, $\nu_i$ have lepton number $+1$ and are referred to as the left-handed neutrino fields while $\nu_i^c$ have lepton number $-1$ and are referred to as the left-handed antineutrino fields. 
Note that while $\nu_i$ couples to weak gauge bosons, $\nu_i^c$ does not.
When it comes to writing down the electromagnetic dipole moments, terms proportional to $\nu_i\sigma_{\mu\nu}\nu_j$ and $\nu^c_i\sigma_{\mu\nu}\nu^c_j$ violate lepton number and are hence forbidden. We are left with
\begin{equation}\label{eq:LagD}
{\cal L}_{\rm D} = \mu^D_{ij} \nu^c_i\sigma_{\mu\nu}\nu_jF^{\mu\nu} + H.c..
\end{equation}
Note that we do not include interactions of the type $\nu_i\sigma_{\mu\nu}\nu^c_j$. These are accounted for since, as already mentioned earlier,  $\nu_i\sigma_{\mu\nu}\nu^c_j=-\nu^c_j\sigma_{\mu\nu}\nu_i$. With this in mind, $\mu^D_{ij}$ define a generic, $N\times N$ complex matrix, parameterized by $N^2$ complex numbers. In the case of three neutrinos, the dipole moment matrix is parameterized by 18 real parameters, $\mu^D_{ij}=|\mu^D_{ij}|e^{i\xi_{ij}}$, $ij=11,12,13,21,22,23,31,32,33$.

There is a useful way to visually compare the Majorana and Dirac dipole moment matrices. In the Majorana case, 
\begin{equation}
{\cal L}_M=\frac{1}{2}\left(
\begin{array}{cccc}
\nu_1 & \nu_2 & \ldots & \nu_N 
\end{array}
\right) \sigma_{\mu\nu}
\left(
\begin{array}{cccc}
0 & \mu_{12} & \ldots & \mu_{1N} \\
-\mu_{12} & 0 & \ldots &  \mu_{2N} \\
\vdots & \vdots & \ddots & \vdots \\
-\mu_{1N} & -\mu_{2N} & \ldots & 0 
\end{array}
\right)
\left(
\begin{array}{c}
\nu_1 \\ \nu_2 \\ \vdots \\  \nu_N 
\end{array}
\right) F^{\mu\nu} + H.c., 
\end{equation}
while in the Dirac case, making use of $\nu_i\sigma_{\mu\nu}\nu^c_j=-\nu^c_j\sigma_{\mu\nu}\nu_i$, we can rewrite Eq.~(\ref{eq:LagD}) in a more ``symmetric'' way, so that it looks very much like the Majorana case:
\begin{equation}
{\cal L}_D=\frac{1}{2}\left(
\begin{array}{cccccc}
\nu^c_1 & \ldots & \nu^c_N & \nu_1 & \ldots & \nu_N 
\end{array}
\right)\sigma_{\mu\nu}
\left(
\begin{array}{cccccc}
0 & \ldots & 0 & \mu^D_{11} & \ldots & \mu^D_{1N} \\ 
\vdots & \ddots & \vdots & \vdots & \ddots & \vdots \\
0 & \ldots & 0 & \mu^D_{N1} & \ldots & \mu^D_{NN} \\ 
-\mu^D_{11} & \ldots & -\mu^D_{N1} & 0 & \ldots & 0 \\ 
\vdots & \ddots & \vdots & \vdots & \ddots & \vdots \\
-\mu^D_{1N} & \ldots & -\mu^D_{NN} & 0 & \ldots & 0  
\end{array}
\right)
\left(
\begin{array}{c}
\nu^c_1 \\ \vdots \\ \nu^c_N \\ \nu_1 \\ \vdots \\  \nu_N 
\end{array}
\right) F^{\mu\nu} + H.c..
\end{equation}
For the same number of neutrino species $N$, the Dirac dipole moment matrix is bigger: $(2N\times2N)$ versus $(N\times N)$. On the other hand, the Dirac dipole moment matrix has a larger fraction of zero entries; in fact, only 1/4 of the entries in the Dirac case are independent and nontrivial. 

It is easy to see that if the number of neutrinos, here defined to be very light neutral fermions, is three, the Dirac case has many more independent dipole moments (18 real parameters) than the Majorana case (6 real parameters). 
Therefore, if the neutrinos are Majorana fermions, the dipole moment matrix can be over-constrained after one obtains 7 independent bits of information. 
On the other hand, 19 independent bits are required in order to over-constrain the dipole moment matrix in the Dirac case.\footnote{Whether these ``bits of information'' are accessible in principle or in practice will be further discussed in the next sections.} 
In Section~\ref{sec:results}, this will translate into the fact that the neutrino electromagnetic dipole moments are less constrained if the neutrinos are Dirac fermions.   

One is tempted to conclude that, by performing enough measurements of the neutrino dipole moments, it is possible to establish the nature of the neutrinos, Majorana fermions versus Dirac fermions. This is not necessarily the case. If the neutrinos are Majorana fermions, one can mimic the Dirac case by adding more neutrino mass eigenstates. For example, by doubling the number of mass eigenstates, the dimensions of the two dipole moment matrices can be made the same. In this case, in fact, there are more independent dipole moments if the neutrinos are Majorana fermions. Concretely, for six Majorana neutrinos, there are 15 complex dipole moments, compared to the 9 complex dipole moments for three Dirac neutrinos. Five Majorana neutrinos, as a matter of fact, are a better ``match'' to three Dirac neutrinos: 10 versus 9 complex parameters. As an aside, the number of independent dipole moments first coincides for 1 Dirac neutrino and 2 Majorana neutrinos, followed by 6 Dirac neutrinos and 9 Majorana neutrinos. The next combinations are 35 Dirac neutrinos versus 50 Majorana neutrinos, followed by 204 Dirac neutrinos versus 289 Majorana neutrinos. We did not find other pairings with less than 1000 Dirac neutrinos.

%%%%%%%%%%%%%%%%%%
\section{Observing Neutrino Electromagnetic Dipole Moments}
\label{sec:observables}
\setcounter{equation}{0}
%%%%%%%%%%%%%%%%%%

A non-zero neutrino electromagnetic  dipole moment modifies elastic neutrino--electron, neutrino--nucleon, and neutrino--nucleus scattering. 
For all processes of interest, the chirality-flipping nature of the magnetic moment, combined with the chirality-conserving nature of the weak interactions and the tiny neutrino masses implies that the contribution from photon-exchange between the neutrino and the charged-fermion of interest will add incoherently to the weak cross section. 
For $\nu_i+e\to \nu_j+e$ elastic scattering, the dipole-moment contribution to the cross section is\begin{eqnarray}
\frac{d \sigma_{ij}}{dE_{R}} =\frac{|\mu_{ji}|^{2}}{\mu_B^2}\frac{\pi \alpha^2}{m_e^2}  \left[\frac{1}{E_{R}}-\frac{1}{E_{\nu}}\right],
\label{eq:nu-e}
\end{eqnarray}
where  $E_\nu$ is the energy of the incoming neutrino, $E_R$ is the electron recoil kinetic energy,  $\alpha$ is the fine-structure constant, $m_e$ is the electron mass, and $\mu_B\equiv e/2m_e$ is the Bohr magneton. 
The signature of the dipole moment in neutrino--electron scattering experiments is an excess of recoil electrons that peaks at small recoil kinetic energies. 
For coherent elastic scattering on nuclei, the cross section is given by Eq.~(\ref{eq:nu-e}) multiplied by $Z^2 F^2(q^2)$, where $Z$ is the atomic number of the target, $F(q^2)$ is the nuclear from factor, and $q^2$ is the four-momentum transfer~\cite{Vogel:1989iv}.

Since neutrino masses are negligibly small and the daughter neutrinos cannot, for all practical purposes, be observed directly or indirectly, $\sigma_{ij}$ is not really an observable. 
Instead, upon detecting the recoil charged particle, one measures $\sigma_i\equiv\sum_j\sigma_{ij}$. 
For neutrino--electron scattering,
\begin{eqnarray}
\frac{d \sigma_{i}}{dE_{R}} =\frac{|\mu_{i}^{\rm eff}|^{2}}{\mu_B^2}\frac{\pi \alpha^2}{m_e^2}  \left[\frac{1}{E_{R}}-\frac{1}{E_{\nu}}\right],
\label{eq:nui-e}
\end{eqnarray}
where 
\begin{equation}
|\mu_i^{\rm eff}|^2\equiv \sum_j|\mu_{ji}|^2,
\label{eq:mu_i_def}
\end{equation}
is the magnitude squared of the effective magnetic moment associated to an incoming $\nu_i$. 
The effective magnetic moments $\mu_i^{\rm eff}$ are directly constrained by solar neutrino experiments since these are best described as incoherent mixtures of the neutrinos with well defined masses, $\nu_1$, $\nu_2$, $\nu_3$, etc.

Neutrinos that are both produced and detected on Earth are best described as coherent linear superpositions of the neutrino mass eigenstates -- the neutrino flavor eigenstates, $\nu_{\alpha}=U_{\alpha i}\nu_i$, $\alpha=e,\mu,\tau$, where $U_{\alpha i}$ are the elements of the unitary lepton mixing matrix. It is simple to define the neutrino electromagnetic moment matrix in the flavor-eigenstate basis. If the neutrinos are Majorana fermions, 
\begin{equation}\label{eq:LagM_f}
{\cal L}_{\rm M} = \frac{1}{2}\mu_{ij} U^*_{\alpha i}\nu_{\alpha}\sigma_{\mu\nu}U^*_{\beta j}\nu_{\beta}F^{\mu\nu} + H.c. = \frac{1}{2}\mu_{\alpha\beta}\nu_{\alpha}\sigma_{\mu\nu}\nu_{\beta}F^{\mu\nu} + H.c.,
\end{equation}
where 
\begin{equation}
\label{eq:mu_ab}
\mu_{\alpha\beta} \equiv U^*_{\alpha i}U^*_{\beta j}\mu_{ij}.
\end{equation}

Instead, if the neutrinos are Dirac fermions, 
\begin{equation}\label{eq:LagD_f}
{\cal L}_{\rm D} = \mu^D_{ij} V^*_{\alpha i}\nu^c_{\alpha}\sigma_{\mu\nu}U^*_{\beta j}\nu_{\beta}F^{\mu\nu} + H.c. = \mu^D_{\alpha\beta}\nu^c_{\alpha}\sigma_{\mu\nu}\nu_{\beta}F^{\mu\nu} + H.c.
\end{equation}
where we introduce a matrix $V$ that relates the left-handed antineutrinos in the mass eigenstate basis to those in the flavor-eigenstate basis. Since there are no weak interactions for the  left-handed antineutrinos, their flavor-eigenstate basis is undetermined and $V_{\alpha i}$ are not physical. We can take advantage of this and choose $V_{\alpha i}=U_{\alpha i}$ so, for Dirac neutrinos, the electromagnetic dipole moment matrix in the flavor-eigenstate basis is also given by Eq.~(\ref{eq:mu_ab}),  with the addition of the superscript $D$ (for Dirac).

Similar to $\sigma_{ij}$, the neutrino dipole contribution to the $\nu_{\alpha}+e\to\nu_{\beta}+e$ cross section $\sigma_{\alpha\beta}$ is proportional to $|\mu_{\beta\alpha}|^2$. Summing over the flavors of the final-state netrinos,
\begin{eqnarray}
\frac{d \sigma_{\alpha}}{dE_{R}} =\frac{|\mu_{\alpha}^{\rm eff}|^{2}}{\mu_B^2}\frac{\pi \alpha^2}{m_e^2}  \left[\frac{1}{E_{R}}-\frac{1}{E_{\nu}}\right],
\label{eq:nua-e}
\end{eqnarray}
where 
\begin{equation}
|\mu_\alpha^{\rm eff}|^2\equiv \sum_\beta|\mu_{\beta\alpha}|^2.
\end{equation}

Note that one is not obliged to work in the flavor-eigenstate basis even when the incoming state is a flavor eigenstate. In the mass-eigenstate basis, the incoming neutrino is a linear superposition of mass eigenstates so the amplitude for $\nu_\alpha \to \nu_i$ is ${\cal A}_{\alpha i}\propto U_{\alpha j}\mu_{ij}$. Summing over all possible final-states (assuming again these are impossible to measure or ``tag'' in either flavor or mass eigenstates) $\sigma_{\alpha}\propto \sum_i|U_{\alpha j}\mu_{ij}|^2$. It is easy to show that $\sum_i|U_{\alpha j}\mu_{ij}|^2=\sum_\beta|\mu_{\beta\alpha}|^2=|\mu_{\alpha}^{\rm eff}|^2$.

There remains the possibility of producing a neutrino flavor-eigenstate $\nu_{\alpha}$ and detecting it via elastic scattering some distance $L$ away from the neutrino source. In this case, the incoming neutrino state is the ``oscillated $\nu_{\alpha}$,'' a different linear superposition of mass-eigenstates (see, for example, \cite{Beacom:1999wx}). Given what is known about the neutrino mass-squared differences, oscillation effects are irrelevant to all Earth-bound experimental constraints of interest.

\section{Summary of Experimental Constraints}
\label{sec:experiments}
\setcounter{equation}{0}
%%%%%%%%%%%%%%%%%%

As discussed earlier, we will concentrate on bounds that come from the scattering of solar neutrinos or Earth-bound (anti)neutrinos.\footnote{We will, in general, use `neutrinos' to refer to neutrinos or antineutrinos.} In the case of Earth-bound neutrinos, different sources have been used in order to search for a nonzero neutrino electromagnetic moment, including neutrinos from nuclear reactors and neutrinos from pion decay. In both cases, the strongest bounds are obtained from precise analyses of neutrino--electron scattering so we will concentrate on those. In the case of neutrinos from pion decay at rest, coherent elastic neutrino--nucleus scattering (CEvNS) data has also been used to search for nonzero neutrino dipole moments. Current estimates, obtained from data made available by the COHERENT Collaboration \cite{COHERENT:2018imc}, are not yet competitive (for recent analyses see \cite{Cadeddu:2020lky,AtzoriCorona:2022qrf}). From the CEvNS measurement in CsI,  constraints down to few~$\times10^{-9}~\mu_B$ can be obtained with $90\%$~C.L, while the future detector materials of the COHERENT experiment, e.g. Ge, can generally perform better by a factor of a few~\cite{Billard:2018jnl,Baxter:2019mcx,Miranda:2019wdy}.

There are also interesting results from the DONUT experiment, which obtains an upper bound of $|\mu^{\rm{eff}}_{\nu_\tau}|<3.9\times10^{-7}~\mu_B$ with $90\%$ C.L.~\cite{DONUT:2001zvi}. It makes use of neutrinos from meson decays in flight, including a nonzero sample of $\nu_{\tau}$-initiated scattering events. We will comment on these bounds in Sec.~\ref{sec:results}. 

One can also obtain constraints down to $|\mu_\nu|\sim{\rm{few}}\times10^{-12}~\mu_B$ from stellar cooling~\cite{Diaz:2019kim}. These constraints are less robust and somewhat model dependent (for an earlier detailed discussion, see, for example, \cite{Heger:2008er}). It has also been argued that new physics can weaken such bounds significantly. For example, in so-called ``chameleon'' models~\cite{Khoury:2003rn}, these bounds are virtually absent. The observation of neutrinos from Supernova 1987A can also be used to constrain the neutrino magnetic moment. Early estimates pointed to $|\mu_\nu|\sim{\rm{few}}\times10^{-13}~\mu_B$~\cite{Lattimer:1988mf,Raffelt:1999gv}. More recently, however, these bounds were called into question~\cite{Bar:2019ifz}, and it was argued that they may not be valid at all. Henceforth, we do not consider indirect astrophysical bounds in our analyses.

Finally, unless otherwise noted, we assume henceforth there are only three light neutrino states. 

%%%%%%%%%%%%%%%%%%
\subsection{Solar Experiments}
%%%%%%%%%%%%%%%%%%

Neutrinos from the Sun arrive at the Earth as incoherent mixtures of the mass eigenstates: $\nu_1$ with probability $P_1$, $\nu_2$ with probability $P_2$, $\nu_3$ with probability $P_3$ (for a recent, detailed overview, see, for example, \cite{Maltoni:2015kca}). Given what is currently known about neutrino-oscillation parameters\footnote{In our analyses, we use the results presented in \cite{Esteban:2020cvm}, NuFIT5.1 (2021). See also \url{http://www.nu-fit.org}. Concretely, we use $\sin^2\theta_{12}=0.304$, $\sin^2\theta_{13}=0.02220$, $\sin^2\theta_{23}=0.573$.}, for all solar neutrino energies, $P_3=|U_{e3}|^2\sim 0.02$, while $P_1$ and $P_2$ depend on the neutrino energy. Here we ignore the impact of the nonzero neutrino magnetic moments on the flavor evolution of the neutrinos inside the Sun.

The measurement of solar neutrinos scattering on electrons, for a fixed neutrino energy, is sensitive to 
\begin{equation}
|\mu|^2_{\rm solar}=P_1|\mu_1^{\rm eff}|^2+P_2|\mu_2^{\rm eff}|^2+P_3|\mu_3^{\rm eff}|^2.
\end{equation}

The best published solar neutrino constraints are from the Borexino experiment. Using solar neutrino data taken in $1291.5$ days during its second phase, Borexino set an upper bound of $|\mu|_{\rm solar}< 2.8\times10^{-11}~\mu_B$ at $90\%$ C.L. for predominantly $^7$Be neutrinos (monochromatic, $E_{\nu}=862$~keV). For $^7$Be neutrino energies, matter effects inside the Sun are small and $P_1=|U_{e1}|^2\sim 0.7$ and $P_2=|U_{e2}|^2\sim 0.3$ to a good approximation. 

The XENON experiments, while searching for dark matter, are also sensitive to neutrinos from the Sun. When it comes to nonzero magnetic moments, the dominant contribution is from $pp$ solar neutrinos (most abundant, lowest energy). For $pp$-solar neutrinos, matter effects inside the Sun are negligible and $P_1=|U_{e1}|^2\sim 0.7$ and $P_2=|U_{e2}|^2\sim 0.3$ is an excellent approximation.  The excess of electron recoil events reported by the XENON1T collaboration~\cite{XENON:2020rca} can be explained by a nonzero neutrino electromagnetic moment ($\mu_\nu=5.7\times10^{-11}$ is the quoted best-fit value~\cite{Brdar:2020quo}). However, the observed excess can also be interpreted as evidence for some unaccounted-for background, e.g., tritium decays~\cite{XENON:2020rca}. Given all the uncertainty, we do not include the XENON1T results in our analysis. 
Furthermore, very recently, first results on the low-energy electron-recoil data of the XENONnT collaboration were made public~\cite{XENON:2022mpc}. 
The XENONnT collaboration reports an upper bound of $|\mu|_{\rm solar}<6.3\times10^{-12}~\mu_B$ (90\% C.L.) that is almost five times stronger than the Borexino upper bound. 
This bound supersedes the XENON1T hint by almost an order of magnitude and is included in our analysis. 

Future dark matter direct-detection experiments will also be sensitive to the $pp$ solar neutrinos. These should be sensitive to effective magnetic moments of order $10^{-12}\mu_B$ \cite{Aalbers:2022dzr}, almost an order of magnitude smaller than the recently reported XENONnT bound.  

There are also constraints from the scattering of $^8$B neutrinos on electrons \cite{Grifols:2004yn}. $^8$B neutrinos have energies between 5~MeV and 10~MeV and are strongly impacted by solar matter effects. For $^8$B neutrinos, $P_1\sim0.1$ and $P_2\sim0.9$, with some energy dependency. Dipole moment constraints from $^8$B neutrinos are not competitive with those from Borexino or XENONnT and will not be included in our results.

%%%%%%%%%%%%%%%%%%
\subsection{Reactor Experiments}
%%%%%%%%%%%%%%%%%%

Nuclear reactors are intense sources of electron antineutrinos. The GEMMA experiment~\cite{Beda:2012zz} sets the strongest bound on the neutrino electromagnetic moment among the reactor neutrino experiments. Using a total of $22,621$ hours of data taking, they set the upper bound $\mu^{\rm{eff}}_{\bar{e}}<2.9\times10^{-11}~\mu_B$ at $90\%$~C.L. (the bar indicates an incoming $\bar{\nu}_e$). 
The TEXONO collaboration also measured elastic neutrino--electron scattering for electron antineutrinos coming from the Kuo-Sheng Nuclear reactor~\cite{TEXONO:2009knm} and constrained $\mu^{\rm{eff}}_{\bar{e}}<2.2\times10^{-10}~\mu_B$ at $90\%$~C.L.. This is an order of magnitude weaker than the GEMMA bound and hence we ignore it here. 
More recently, the CONUS collaboration, using candidate neutrino--electron scattering events, also reported a bound on the the effective electron antineutrino magnetic moment, $\mu^{\rm{eff}}_{\bar{e}}<7.5\times10^{-11}~\mu_B$ at $90\%$~C.L. \cite{CONUS:2022qbb}. 
Since it is two and half times weaker than the published GEMMA bounds, we do not include the CONUS constraints in our analyses. 

%%%%%%%%%%%%%%%%%%
\subsection{Accelerator Experiments}
%%%%%%%%%%%%%%%%%%
%cite[0101039]
The LSND experiment measured neutrino--electron scattering using neutrinos produced in $\pi^+$ and $\mu^+$ decay at rest~\cite{LSND:2001akn}. Pion decay produces mostly $\nu_{\mu}$ while muon decay produces both $\nu_e$ and $\bar{\nu}_{\mu}$. LSND data are analyzed and the collaboration reports a constraint on a mixture of $|\mu^{\rm{eff}}_{e}|^2$ and $|\mu^{\rm{eff}}_{\mu}|^2$:
$|\mu^{\rm{eff}}_{e}|^2+2.4|\mu^{\rm{eff}}_{\mu}|^2<1.1\times10^{-18}~\mu_B^2$ at $90\%$~C.L.~\cite{LSND:2001akn}. They assume $|\mu^{\rm{eff}}_{\mu}|^2=|\mu^{\rm{eff}}_{\bar{\mu}}|^2$.

In the future, the DUNE experiment is expected to be sensitive to $|\mu^{\rm{eff}}_{\mu}|>3.2\times10^{-10}~\mu_B$ at $90\%$~C.L. after seven years data taking in both the neutrino and antineutrino modes~\cite{Mathur:2021trm}. Because of the GeV energy range of DUNE and the dependence of the electromagnetic cross section on the inverse of the neutrino energy, DUNE is not the best place to get competitive constraint on $\mu^{\rm{eff}}_{\mu}$, despite its unprecedented neutrino flux and large detector mass. The J-PARC Sterile Neutrino Search at J-PARC Spallation Neutron Source (JSNS$^2$) experiment \cite{Ajimura:2017fld}, including its proposed upgrade \cite{Ajimura:2020qni}, might ultimately have better sensitivity since it makes use of neutrinos from meson and muon decay at rest, similar to LSND. Finally, as already discussed, future measurements of CEvNS and neutrino--electron scattering using neutrinos from pion decay at rest may ultimately provide better sensitivity to $|\mu^{\rm{eff}}_{\mu}|$. 

To illustrate the impact a measurement of $|\mu_\mu^{\rm eff}|$ could have on the experimental landscape, in Sec.~\ref{sec:results} we will assume that results from a future experiment sensitive to $|\mu^{\rm{eff}}_{\mu}|>2\times 10^{-11}\mu_B$ are available. 
This sensitivity is comparable to that of Borexino and does not compete with expectations from future solar experiments. 
Nonetheless, we will argue that the impact of such an experiment may be, under the right circumstances, very significant. 

%%%%%%%%%%%%%%%%%%
\subsection{Statistical Treatment of Experimental Constraints.}
%%%%%%%%%%%%%%%%%%

All experiments report upper bounds on some effective electromagnetic moment $|\mu^{\rm eff}|_{\rm exp}$ (in general a different effective magnetic moment for each experiment of interest). When computing upper bounds on the different $|\mu_{ij}|$, presented and discussed in Sec.~\ref{sec:results}, we treat these upper bounds as quadratic $\chi^2$ functions of $|\mu_{\rm eff}|^2$ and assume the best-fit values associate to all experimental results are equal to zero:
\begin{equation}
\chi^2_{\rm exp} = \frac{(|\mu^{\rm eff}|^2)^2}{\sigma^2_{\rm exp}},
\end{equation}   
where $\sigma_{\rm exp}$ is extracted from the reported 90\% C.L. upper bounds quoted by the different collaborations, $\mu_{90\%}^{\rm eff}$:
\begin{equation}
\sigma_{\rm exp}^2 = \frac{(\mu_{90\%}^{\rm eff})^4}{2.7}.
\end{equation}
The reason for this assumption is that the number of dipole-moment-mediated events at any experiment is linearly proportional to $|\mu^{\rm eff}|^2$ as can be seen, for example, in Eqs.~(\ref{eq:nui-e}) and (\ref{eq:nua-e}). Note that, traditionally, one quotes upper bounds on $|\mu^{\rm eff}|$.
In order to combine results from different experiments, we assume the total $\chi^2$ to be sum of all the relevant $\chi^2_{\rm exp}$. 
While this may be an oversimplification, as we are assuming the best fits to be null and neglecting correlations (e.g. in solar neutrino fluxes), we find this approach to be suitable to make our point on the interplay between magnetic moment measurements and the nature of neutrinos.
%This is the object we use to compute upper limits and bounds on subsets of the parameter space. 

%%%%%%%%%%%%%%%%%%
\section{Results: Present and Future}
\label{sec:results}
\setcounter{equation}{0}
\setcounter{footnote}{0}
%%%%%%%%%%%%%%%%%%

Here we present and discuss the current constraints on all $|\mu_{ij}|$ for both Majorana and Dirac neutrinos. We present all results in the neutrino mass-eigenstate basis; when convenient, we make use of the flavor-eigenstate basis in order to discuss specific results. Our ultimate goal is to combine all constraints from the different neutrino sources and experiments and discuss the impact of future experimental efforts. We comment on individual constraints when it is illuminating. Comparisons between Dirac neutrinos and Majorana neutrinos are presented in the `Dirac Neutrinos' subsection.  

All upper bounds and exclusion curves are quoted at 90\% C.L., for the relevant number of degrees of freedom.

\subsection{Majorana Neutrinos}

If neutrinos are Majorana fermions, assuming there are no new light fermions, there are three independent complex neutrino electromagnetic dipole moments: $\mu_{12}, \mu_{13}, \mu_{23}$. We will concentrate on the existing constraints on $|\mu_{12}|, |\mu_{13}|, |\mu_{23}|$, keeping in mind the complex phases in $\mu_{12}, \mu_{13}, \mu_{23}$ are unconstrained. Unless otherwise noted, when presenting constraints on $|\mu_{12}|, |\mu_{13}|, |\mu_{23}|$, we marginalize over all unreported parameters. 

\begin{figure}[t]
\begin{center}
\includegraphics[angle=0,width=1\textwidth]{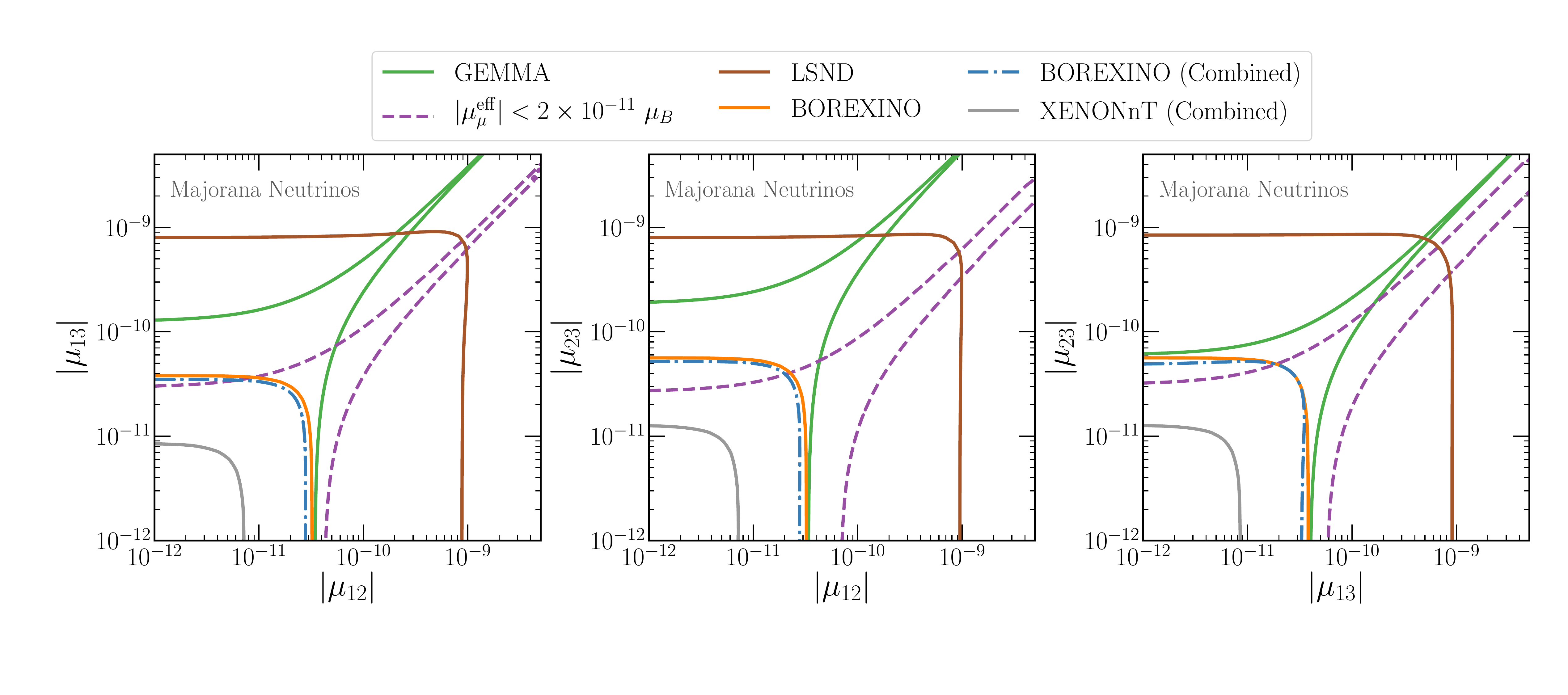}
    \vspace*{-1.2cm}
    \caption{Majorana neutrinos. 90\% C.L. allowed regions in the 
    $|\mu_{12}|\times|\mu_{13}|$-plane (left), 
    $|\mu_{12}|\times|\mu_{23}|$-plane (center), $|\mu_{13}|\times|\mu_{23}|$-plane (right), 
    extracted from different subsets of existing and hypothetical future data. `BOREXINO (Combined)' stands for data from Borexino, GEMMA, and LSND. `XENONnT (Combined)' stands for data from XENONnT, Borexino, GEMMA, and LSND. `$|\mu^{\rm eff}_{\mu}|<2\times 10^{-11}~\mu_B$' stands for data from a future experiment that constraints $|\mu^{\rm eff}_{\mu}|<2\times 10^{-11}~\mu_B$. See Section~\ref{sec:experiments} for details.}   
    \label{fig:maj_magplots}
\end{center}
\end{figure}

As discussed in Sec.~\ref{sec:observables}, to a good approximation, the solar neutrino experiments of interest are sensitive to 
\begin{equation}
|\mu|^2_{\rm solar}=|U_{e1}|^2\left(|\mu_{12}|^2+|\mu_{13}|^2\right)+
|U_{e2}|^2\left(|\mu_{12}|^2+|\mu_{23}|^2\right) + 
|U_{e3}|^2\left(|\mu_{13}|^2+|\mu_{23}|^2\right),
\label{eq:solar_maj}
\end{equation}
and hence insensitive to the relative phases among the different dipole moments. Constraints from Borexino and XENONnT in the different $|\mu_{ij}|\times |\mu_{jk}|$ planes ($i,j,k=1,2,3$) are depicted in Fig.~\ref{fig:maj_magplots} (orange and grey lines, respectively). Since all terms in Eq.~(\ref{eq:solar_maj}) are positive-definite, it is possible to marginalize over all-but-one of the elements of the electromagnetic dipole matrix and constrain each $|\mu_{ij}|$ independently. The 90\% C.L. upper bounds we obtain from the Borexino and XENONnT bounds are listed in Table~\ref{tab:MajBnds}. Throughout, we kept the neutrino oscillation parameters fixed at their best-fit values, except for the CP-odd parameter $\delta_{\rm CP}$, which we allow to float in the fits.  Note that the CP-odd phase is irrelevant for the solar neutrino constraints. Had we allowed the mixing angles to also float in the fits, we would have obtained slightly weaker bounds (roughly five to ten percent), given the current uncertainties on the relevant mixing parameters. 
%%%
%%%

Table~\ref{tab:MajBnds} reveals that the constraints from solar data on $|\mu_{12}|, |\mu_{13}|, |\mu_{23}|$ are relatively similar, within less than a factor of two (a factor a little over three for $|\mu_{ij}|^2$). The reason is that, for Majorana neutrinos, $\mu_{ij}=-\mu_{ji}$. Even though $|U_{e3}|^2\ll |U_{e1}^2|,|U_{e2}|^2$, the coefficients behind the different $|\mu_{ij}|^2$ are relatively similar, ranging from $|U_{e2}|^2+|U_{e3}|^2\sim 0.3$ to $|U_{e1}|^2+|U_{e2}|^2\sim 1$.
%%%
\begin{table}
\begin{center}
\caption{90\% C.L. upper bounds on the magnitudes of the different entries of the neutrino electromagnetic moment matrix, for Majorana neutrinos, extracted from different subsets of existing and hypothetical future data. `Future $\nu_{\mu}$' stands for a future experiment capable of constraining $|\mu^{\rm eff}_{\mu}|<2\times 10^{-11}~\mu_B$. See Section~\ref{sec:experiments} for details.
\label{tab:MajBnds}\\}
\begin{tabular}{l|c|c|c}\toprule
Experiment & $|\mu_{12}|~(10^{-11}\mu_B)$ & $|\mu_{13}|~(10^{-11}\mu_B)$ & $|\mu_{23}|~(10^{-11}\mu_B)$ \\
\hline\hline
LSND  & $90$ & $84$& $79$ \\
Borexino & $2.8$ & $3.3$  & $5.0$ \\ 
Borexino \& LSND \& GEMMA & $2.4$ & $3.0$ & $4.4$ \\ 
XENONnT & $0.64$ & $0.75$ &$1.1$  \\
All Combined  & $0.64$ & $0.75$ & $1.1$\\
All Combined \& Future $\nu_{\mu}$ & $0.64$ & $0.75$ & $1.1$\\
\hline\hline
%\addlinespace
%Combined & 2.47 \ZT{2.42} & 3.08  \ZT{3.06} & 4.55  \ZT{4.55} \\ 
%\bottomrule
\end{tabular}
\end{center}
\end{table}
%%%

The situation is different for experiments that constrain $|\mu_{\alpha}^{\rm eff}|^2$, $\alpha=e,\mu,\tau$, including reactor experiments. Constraints from GEMMA on $|\mu_{\bar{e}}^{\rm eff}|^2$ translate into the green contours in Fig.~\ref{fig:maj_magplots} while the sensitivity of a hypothetical future experiment that can see a nonzero $|\mu_{\mu}^{\rm eff}|^2$ if it is larger than $2\times 10^{-11}\mu_B$ is depicted in purple (dashed line). In both these cases,  there is a clear ``flat direction'' in the different $|\mu_{ij}|\times |\mu_{jk}|$-planes \cite{Canas:2016kfy,Miranda:2019wdy}. This implies, for example, one cannot obtain bounds on any of the $|\mu_{ij}|$ that is independent from the other parameters that define the dipole moment matrix. 

The reason for the flat direction is easy to understand. In the flavor-eigenstate basis,
\begin{equation}
|\mu^{\rm eff}_e|^2 = |\mu_{e\mu}|^2 + |\mu_{e\tau}|^2.
\end{equation}
It is easy to see that $|\mu^{\rm eff}_e|^2$ depends only on the magnitudes of two out of the three  $\mu_{\alpha\beta}$ ($\alpha,\beta=e,\mu,\tau$); it does not depend on $\mu_{\mu\tau}$ at all. Since the three $\mu_{\alpha\beta}$ (and $\mu_{ij}$) are, in general, independent, there is a combination of $|\mu_{\ij}|$ -- indeed, $\mu_{\mu\tau}$ -- that remains unconstrained. This translates into the cuspy contours observed in Fig.~\ref{fig:maj_magplots}. The same argument holds for $|\mu^{\rm eff}_\mu|^2$, $|\mu^{\rm eff}_\tau|^2$.

Flat directions are lifted if one combines constraints on different $|\mu^{\rm eff}_\alpha|^2$. Bounds from the LSND experiments, depicted in brown in Fig.~\ref{fig:maj_magplots}, illustrate this, since, as discussed in Section~\ref{sec:experiments}, LSND constrains a weighted sum of $|\mu^{\rm eff}_\mu|^2$ and $|\mu^{\rm eff}_e|^2$. For this reason, we can compute the LSND bounds on the different $|\mu_{ij}|$ after one marginalizes over all other dipole moment observables. These are listed in Table~\ref{tab:MajBnds}. The LSND bounds are much weaker than the solar bounds. 

Combinations of solar data with reactor or accelerator data are also free from flat directions and one can obtain constraints on all $\mu_{ij}$, marginalizing over all other dipole moment observables, from all current experiments combined. These are listed in Table~\ref{tab:MajBnds} and depicted in Fig.~\ref{fig:maj_magplots} (dot-dashed blue contour for Borexino combined with LSND and GEMMA, grey for XENONnT combined with all other existing data). The bounds from solar experiments dominate those from Earth-bound experiments. The XENONnT constraints are strong enough that the impact of combining them with all other data is negligible. We also combine all existing constraints with a future experiment that excludes, at the 90\% C.L., $|\mu_{\mu}^{\rm eff}|<2\times 10^{-11}\mu_B$. These are listed in Table~\ref{tab:MajBnds}. The impact of the future experiment is negligible relative to that of XENONnT. 

More generally, if the neutrinos are Majorana fermions and there are no extra neutrino degrees of freedom, expectations are that next-generation experiment sensitive to $|\mu^{\rm eff}_\mu|^2>2\times 10^{-11}\mu_B$ will not see the effects of nonzero neutrino electromagnetic moments. The solar bounds preclude it. This is depicted in Fig.~\ref{fig:maj_magplots}. The sensitivity region of the future $|\mu^{\rm eff}_\mu|^2$ is well inside the region of parameters space ruled out by the XENONnT experiment. 

\subsection{Dirac Neutrinos}

If neutrinos are Dirac fermions, assuming there are no new light fermions, there are nine independent complex neutrino electromagnetic dipole moments: $\mu^D_{11},\mu^D_{12}, \mu^D_{13}, \mu^D_{21}, \mu^D_{22},\mu^D_{23},\mu^D_{31},\mu^D_{32},\mu^D_{33}$. Like in the Majorana neutrino case, we will concentrate on the existing constraints on the magnitudes of the different electromagnetic moments $|\mu^D_{ij}|$ ($i,j=1,2,3$), keeping in mind the complex phases of the different $\mu^D_{ij}$ are unconstrained. Unless otherwise noted, when presenting constraints, we marginalize over all unreported parameters. 

\begin{figure}[t]
\begin{center}
\includegraphics[angle=0,width=1\textwidth]{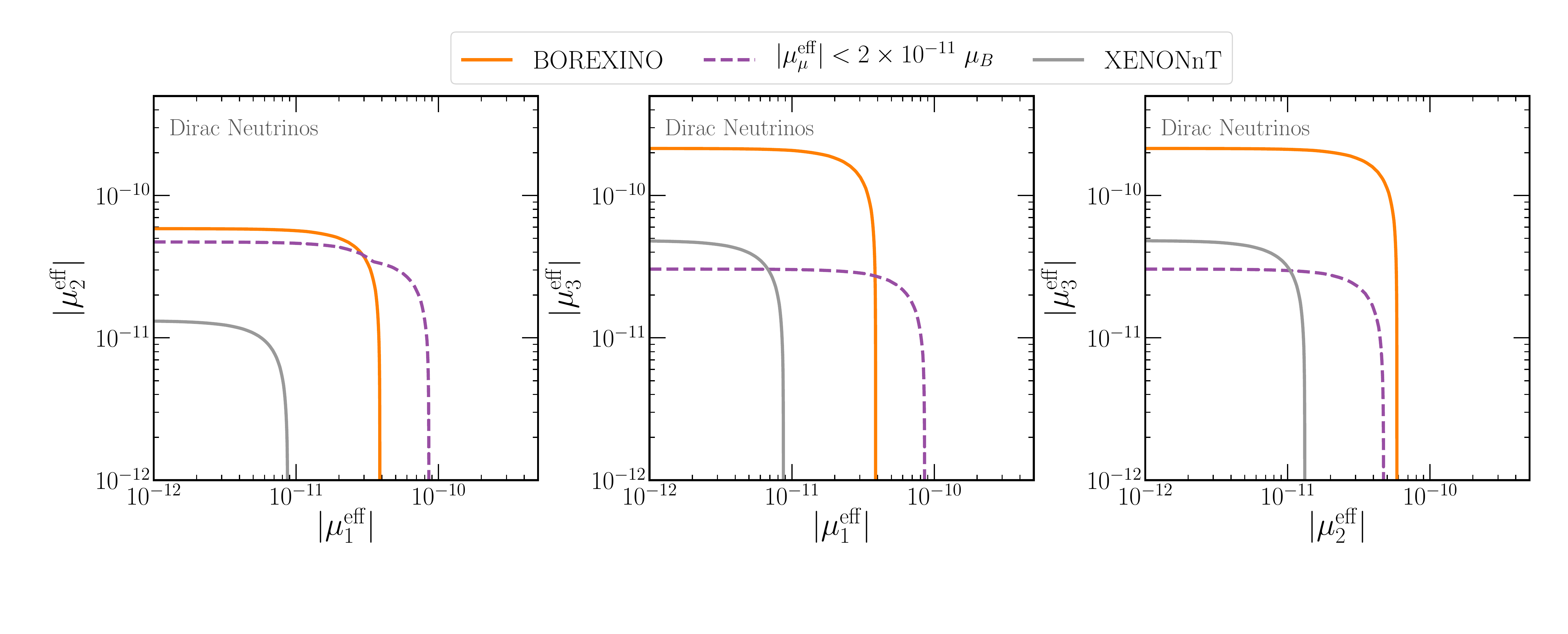}
    \vspace*{-1.2cm}
    \caption{Dirac neutrinos. 90\% C.L. allowed regions in the 
    $|\mu^{\rm eff}_{1}|\times|\mu^{\rm eff}_{2}|$-plane (left), 
    $|\mu^{\rm eff}_{1}|\times|\mu^{\rm eff}_{3}|$-plane (center), $|\mu^{\rm eff}_{2}|\times|\mu^{\rm eff}_{3}|$-plane (right), 
    extracted from different subsets of existing and hypothetical future data. 
    `$|\mu^{\rm eff}_{\mu}|<2\times 10^{-11}~\mu_B$' stands for data from a future experiment that constraints $|\mu^{\rm eff}_{\mu}|<2\times 10^{-11}~\mu_B$.
    See Section~\ref{sec:experiments} for details. 
    In the case of the future experiment sensitive to $|\mu^{\rm eff}_{\mu}|$ (dashed, purple curves), we assume only $\mu^D_{11}, \mu^D_{22}, \mu^D_{33}\neq 0$.
    }   
    \label{fig:dirac_magplots}
\end{center}
\end{figure}

As discussed in Sec.~\ref{sec:observables}, to a good approximation, the solar neutrino experiments of interest are sensitive to 
\begin{eqnarray}
|\mu|^2_{\rm solar}&=& |U_{e1}|^2|\mu_{1}^{\rm eff}|^2 + |U_{e2}|^2|\mu_{2}^{\rm eff}|^2 + |U_{e3}|^2|\mu_{3}^{\rm eff}|^2 \label{eq:solar_dir}, \\
&=&|U_{e1}|^2\left(|\mu_{11}^D|^2+|\mu_{21}^D|^2+|\mu_{31}^D|^2\right) +  |U_{e2}|^2\left(|\mu_{12}^D|^2+|\mu_{22}^D|^2+|\mu_{32}^D|^2\right)+ \nonumber \\
&& + |U_{e3}|^2\left(|\mu_{13}^D|^2+|\mu_{33}^D|^2+|\mu_{33}^D|^2\right),
\end{eqnarray}
and hence insensitive to the relative phases among the different dipole moments. 
While all nine $|\mu^D_{ij}|$ are constrained by solar data, it is clear that the bounds are correlated. 
After marginalizing over all other $|\mu^D_{ij}|$, the bounds on, for example, $|\mu^D_{11}|$ and $|\mu^D_{21}|$ are identical. 
Hence, as far as solar data are concerned, it is sufficient to extract bounds on $|\mu_i^{\rm eff}|$, $i=1,2,3$, defined in Eq.~(\ref{eq:mu_i_def}); these apply to all $|\mu^D_{ij}|$ (for fixed $i$, $j=1,2,3$).  90\% C.L. 
Constraints from Borexino and XENONnT in the different $|\mu^{\rm eff}_{i}|\times |\mu^{\rm eff}_{j}|$ planes ($i,j=1,2,3$) are depicted in Fig.~\ref{fig:dirac_magplots} in orange and grey, respectively. Since all terms in Eq.~(\ref{eq:solar_dir}) are positive-definite, it is possible to marginalize over all-but-one of the effective electromagnetic dipole moments and constrain each $|\mu^{\rm eff}_{i}|$ independently. 
The 90\% C.L. upper bounds we obtain from the Borexino and XENONnT bounds are listed in Table~\ref{tab:DirBnds}. 
Throughout, we kept the neutrino oscillation parameters fixed at their best-fit values, except for the CP-odd parameter $\delta_{\rm CP}$, which we allow to float in the fits.  Note that the CP-odd phase is irrelevant for the solar neutrino constraints. Had we allowed the mixing angles to also float in the fits, we would have obtained slightly weaker bounds (roughly five to ten percent), given the current uncertainties on the relevant mixing parameters. 

%
%
%%%
%\begin{table}
%\begin{center}
%\caption{Neutrino electromagnetic moment matrix element bounds in the mass-eigenstate basis for a Dirac neutrinos. \textcolor{blue}{[GJS: I'm using the BORX + GEMMA results for the Combined results. This seems to hold true for the plots, and the combined analysis didn't find lower minima.]} All magnetic-moment values shown are in units of $10^{-11}\mu_B$. \label{tab:DirBnds}\\}
%\begin{tabular}{lccccccccc}\toprule
%Experiment & $|\mu_{12}|$ & $|\mu_{13}|$ & $|\mu_{23}|$ & $|\mu_{21}|$ & $|\mu_{31}|$ & $|\mu_{32}|$ & $|\mu_{11}|$ & $|\mu_{22}|$ & $|\mu_{33}|$\\
%\midrule
%GEMMA & --- & --- & --- & --- & --- & --- & --- & --- & --- \\ 
%LSND  & --- & --- & --- & --- & --- & --- & --- & --- & --- \\
%BORX & 5.14 & 18.8 & 18.8 & 3.40 & 3.40 & 5.14 & 3.40 & 5.14 & 18.8 \\ 
%\addlinespace
% B + G & 4.83332 & 17.67969 & 17.67573 & 3.20178 & 3.20178 & 4.83431 & 3.20178 & 4.83233 & 17.66682 \\
%BORX+GEMMA & 4.833 & 17.679 & 17.675 & 3.20 & 3.20 & 4.834 & 3.20 & 4.832 & 17.666 \\
%BORX+LSND & 5.14 & 18.8 & 18.8 & 3.40 & 3.40 & 5.14 & 3.40 & 5.14 & 18.8 \\
%\addlinespace
%Combined* & 4.833 & 17.679 & 17.675 & 3.20 & 3.20 & 4.834 & 3.20 & 4.832 & 17.666 \\ 
%\bottomrule
%\end{tabular}
%\end{center}
%\end{table}
%%%
\begin{table}
\begin{center}
\caption{90\% C.L. upper bounds on the magnitudes of the different entries of the neutrino electromagnetic moment matrix, for Dirac neutrinos, extracted from different subsets of existing and hypothetical future data. `Future $\nu_{\mu}$' stands for a future experiment capable of constraining $|\mu^{\rm eff}_{\mu}|<2\times 10^{-11}~\mu_B$. See Section~\ref{sec:experiments} for details. \label{tab:DirBnds}\\}
%\begin{tabular}{l|c|c|c}\toprule
%\footnotesize Experiment & \footnotesize  $|\mu^D_{11}|, |\mu^D_{21}|, |\mu^D_{31}|~(10^{-11}\mu_B)$ & \footnotesize $|\mu^D_{12}|, |\mu^D_{22}|, |\mu^D_{32}|~(10^{-11}\mu_B)$ & \footnotesize $|\mu^D_{13}|, |\mu^D_{23}|, |\mu^D_{33}|~(10^{-11}\mu_B)$ \\
%\hline\hline
%\footnotesize Borexino & $3.4$ & $5.1$   & $19$  \\ 
%\footnotesize Borexino \& GEMMA \& LSND & $3.2$ & $4.8$ &$18$ \\ 
%\footnotesize XENONnT & $0.76$ & $1.2$ &$4.2$ \\
%\footnotesize All Combined  & $0.76$ & $1.2$ & $4.2$ \\
%\footnotesize All Combined \& Future  $\nu_\mu$  & $0.76$ & $1.2$  & $2.8$ \\ 
%\hline
%\hline
%\end{tabular}
\begin{tabular}{l|c|c|c}\toprule
&\multicolumn{3}{c}{$|\mu_{ij}^D|~(10^{-11}\mu_B)$}\\ \hline
 Experiment &  ~$ij=$ 11, 21, 31~ &  ~$ij=$ 12, 22, 32~ &  ~$ij=$ 13, 23, 33~\\
\hline\hline
 Borexino & $3.4$ & $5.1$   & $19$  \\ 
 Borexino \& GEMMA \& LSND & $3.2$ & $4.8$ &$18$ \\ 
 XENONnT & $0.76$ & $1.2$ &$4.2$ \\
 All Combined  & $0.76$ & $1.2$ & $4.2$ \\
 All Combined \& Future  $\nu_\mu$  & $0.76$ & $1.2$  & $2.8$ \\ 
\hline
\hline
\end{tabular}
\end{center}
\end{table}
%%%

Table~\ref{tab:DirBnds} reveals that $|\mu^{\rm eff}_{3}|$ is significantly less constrained -- one order of magnitude -- by solar data than $|\mu^{\rm eff}_{1,2}|$. The reason is that, for Dirac neutrinos, the different $|\mu^{\rm eff}_{i}|$ are independent and $|U_{e3}|^2\ll |U_{e1}|^2,|U_{e2}|^2$. This is to be contrasted to the Majorana case, where all independent $|\mu_{ij}|$ are similarly constrained by solar data. In the Dirac case, if $|U_{e3}|^2$ were zero, the bound on $|\mu^{\rm eff}_{3}|$ would disappear. In the Majorana case, the solar bounds presented in Table~\ref{tab:MajBnds} would be almost identical to what one would have obtained if $|U_{e3}|^2$ were zero.

Experimental results that translate into an upper bound on a single $|\mu^{\rm eff}_{\alpha}|$, $\alpha=e,\mu,\tau$, do not translate into bounds on individual $\mu^D_{ij}$, similar to the Majorana case. Also here, there are flat directions, i.e., linear combinations of $|\mu^D_{ij}|^2$ that are unconstrained. In fact, in the Dirac case, there are many more flat directions relative to the Majorana case. This is simplest to see in the flavor-eigenstate basis. For example,  
\begin{equation}
|\mu^{\rm eff}_{e}|^2 = |\mu^D_{ee}|^2 + |\mu^D_{\mu e}|^2 + |\mu^D_{\tau e}|^2,
\end{equation}
clearly independent from six of the nine $|\mu^D_{\alpha\beta}|$, $\alpha,\beta=e,\mu,\tau$. 

Unlike the Majorana case, constraints from LSND are also plagued by flat directions in the Dirac case. Using the flavor-eigenstate basis, the effective dipole moment constrained by LSND is independent from $|\mu_{e\tau}|^2, |\mu_{\mu\tau}|^2,|\mu_{\tau\tau}|^2$. In the case of Dirac neutrinos, a collection of Earth-bound experiments capable of fully constraining all independent $|\mu^D_{ij}|$ should also include a $\nu_{\tau}$-initiated scattering sample.\footnote{Another option is an ``oscillated'' scattering sample, i.e, a well-defined flavor eigenstate detected via the electromagnetic dipole-moment interaction a long distance $L$ away.} For example, data from GEMMA, LSND, \emph{and} DONUT can constrain all $\mu_{ij}^D$, independent from exact flat directions. 

Combinations of solar data with those from Earth-bound experiments are not, of course, plagued by flat directions. Furthermore, Earth-bound experiments provide information on the $|\mu^D_{ij}|$ beyond $|\mu^{\rm eff}_{i}|^2$. We return to these momentarily but, for now, it is enough to state that all such combinations still translate into identical bounds on the elements of the ``triplets'' $(|\mu^D_{11}|, |\mu^D_{21}|, |\mu^D_{31}|)$, $(|\mu^D_{12}|$, $|\mu^D_{22}|, |\mu^D_{32}|)$, or $(|\mu^D_{13}|, |\mu^D_{23}|, |\mu^D_{33}|)$. In practice, given that constraints from Borexino and, especially, XENONnT are stronger than those from nuclear reactor and accelerator neutrinos, the consequences of adding, to the solar data, the Earth-bound data, are quantitatively quite small. Combined results are listed in Table~\ref{tab:DirBnds}. As in the Majorana case, the XENONnT constraints are strong enough that the impact of combining them with all other data is negligible. 

Future data could, in principle, lead to a less trivial picture and more information. Constraints from an experiment that rules out, at the 90\% C.L., $|\mu^{\rm eff}_{\mu}|^2>2\times 10^{-11}\mu_B$, combined  with current XENONnT data, are also listed in Table~\ref{tab:DirBnds}. While the impact on the $|\mu_{i1}|$ and $|\mu_{i2}|$ ($i=1,2,3$) elements is negligible, the impact on the $|\mu_{i3}|$ elements is quite significant. This is due to the fact that $|U_{e3}|\ll |U_{\mu 3}|$. More important than placing more stringent bounds, if the neutrinos are Dirac fermions, a future experiment more sensitive to $|\mu^{\rm eff}_{\mu}|^2$ than LSND may potentially observe the effect of a nonzero neutrino electromagnetic even if there are no extra neutrino states. Fig.~\ref{fig:dirac_magplots} depicts the sensitivity of a hypothetical future experiment that can see a nonzero $|\mu_{\mu}^{\rm eff}|^2$ if it is larger than $2\times 10^{-11}\mu_B$ in purple (dashed), assuming only the diagonal $\mu^D_{ij}$ are nonzero.\footnote{The flat directions, discussed earlier, are the reason for restricting here the 18-dimensional parameter space to this much smaller subspace. Otherwise, defining the sensitivity of the future $\nu_{\mu}$ experiment would be both cumbersome and opaque.}
The figure reveals that the sensitivity of such an experiment reaches beyond current constraints on $|\mu^{\rm eff}_3|$. This is qualitatively different from what was observed in the Majorana case, Fig.~\ref{fig:maj_magplots}. There, an experiment sensitive to $|\mu_{\mu}^{\rm eff}|^2>2\times 10^{-11}\mu_B$ is unable to make a discovery unless there are light fermionic states other than the known neutrinos. 

We now turn to the details of the experimental sensitivity of scattering experiments to $|\mu^D_{ij}|$.
In the mass-eigenstate basis, reorganizing the terms in the summations, indicated here explicitly,
\begin{equation}
|\mu_{\alpha}^{\rm eff}|^2 = \sum_{j,k}\left(\sum_i U_{\alpha j}U^*_{\alpha k}\mu^D_{ij}\mu^{D*}_{ik}\right) \equiv \sum_{j,k}A^{\alpha}_{jk}S_{jk}.
\label{eq:mua_def}
\end{equation}
Here, $A^{\alpha}_{jk}\equiv U_{\alpha j}U^*_{\alpha k}$ depend only on the elements of the mixing matrix,\footnote{This discussion can be trivially generalized to the ``oscillated $\nu_{\alpha}$.''} independent from the values of the electromagnetic moments. Instead, $S_{jk}\equiv\sum_i\mu^D_{ij}\mu^{D*}_{ik}$ depend only on (products of) the electromagnetic dipole moments. 
%The second equality in Eq.~(\ref{eq:mua_def}) is, in spite of appearances, real, since $A^{\alpha}_{jk}=(A^{\alpha}_{jk})^*$ and $S_{jk}=S_{kj}^*$ and $|\mu_{\alpha}^{\rm eff}|^2\propto\Re(A^{\alpha}S)$. 
Eq.~(\ref{eq:mua_def}) also holds for incoming neutrinos that are incoherent superpositions of the mass eigenstates, like the solar neutrinos. In theses cases, $A_{jk}=P_j\delta_{jk}$, where $P_j$ is the probability that an incoming $\nu_j$ ``hits'' the target of interest.

Any combination of measurements of $|\mu_{\alpha}^{\rm eff}|^2$ and $|\mu_i^{\rm eff}|^2$ is capable of measuring, or constraining, at most, the different $S_{jk}$, not necessarily the nine individual $|\mu^D_{ij}|$. When it comes to information on the different $|\mu^D_{ij}|$, this has interesting consequences related to the fact that $S_{jk}\equiv\sum_i\mu^D_{ij}\mu^{D*}_{ik}$, for fixed $j,k$, is invariant under relabeling the ``$i$'' index. In other words, all permutations of the ``$i$'' indices lead to the same $S_{jk}$ and hence the same $\mu^{\rm eff}_{\alpha}$ for all $\alpha$.

Some consequences of this symmetry are important for discussing upper bounds on the different $\mu^D_{ij}$. For example, if one marginalizes any collection of upper bounds (expressed, for concreteness, as a $\chi^2$-function) relative to all but one $|\mu^D_{ij}|$ one will obtain the same reduced $\chi^2$ for all values of $i$ and fixed value of $j$. Hence, the upper bounds one obtains for $|\mu^D_{1j}|,|\mu^D_{2j}|,|\mu^D_{3j}|$ are the same for each value of $j$. This is trivial to see in the solar data, as discussed earlier.  

Generalizing, if the same collection of bounds is marginalized over all but a specific pair $|\mu^D_{ij}|,|\mu^D_{i'k}|$, the same reduced $\chi^2$ is expected for all pairs $i,i'$ related by different permutations of the $i,i'$ indices. For $i=i'$, $j=k$, for example, we recover the result mentioned above, that the bounds on $|\mu^D_{1j}|,|\mu^D_{2j}|,|\mu^D_{3j}|$ are the same for each value of $j$. For $i=i'$, $j\neq k$, constraints in the $|\mu^D_{1j}|\times|\mu^D_{1k}|$, $|\mu^D_{2j}|\times|\mu^D_{2k}|$, $|\mu^D_{3j}|\times|\mu^D_{3k}|$ planes are all the same. Finally, for $i\neq i'$and fixed $j,k$, constraints in the $|\mu^D_{1j}|\times|\mu^D_{2k}|$, $|\mu^D_{2j}|\times|\mu^D_{1k}|$, $|\mu^D_{1j}|\times|\mu^D_{3k}|$, $|\mu^D_{3j}|\times|\mu^D_{1k}|$, $|\mu^D_{2j}|\times|\mu^D_{3k}|$, $|\mu^D_{3j}|\times|\mu^D_{2k}|$ planes are all the same. When $j=k$, only half of these are independent since, for example, the $|\mu^D_{12}|\times|\mu^D_{32}|$ and $|\mu^D_{32}|\times|\mu^D_{12}|$ planes are the same.

Therefore, when it comes to depicting constraints in the planes defined by pairs of $\mu^D_{ij}$, instead of 36 independent such constraints, all accessible information can be depicted in 9 independent planes. Explicitely, these are (the `$=$' signs here mean that, in all the ``equal'' planes the constraints are identical.)
\begin{itemize}
\item $|\mu^D_{ij}|\times|\mu^D_{i'j}|$, $j=1,2,3$:
\begin{eqnarray}
& |\mu^D_{11}|\times|\mu^D_{21}| = |\mu^D_{11}|\times|\mu^D_{31}| = |\mu^D_{21}|\times|\mu^D_{31}|, \nonumber \\
& |\mu^D_{12}|\times|\mu^D_{22}| = |\mu^D_{12}|\times|\mu^D_{32}| = |\mu^D_{22}|\times|\mu^D_{32}|,  \nonumber  \\
& |\mu^D_{13}|\times|\mu^D_{23}| = |\mu^D_{13}|\times|\mu^D_{33}| = |\mu^D_{23}|\times|\mu^D_{33}|. \nonumber 
\end{eqnarray}
\item $|\mu^D_{ij}|\times|\mu^D_{ik}|$, $j\neq k$. The distinguishable $\{j,k\}$ pairs are $\{1,2\}, \{1,3\}, \{2,3\}$:
\begin{eqnarray}
& |\mu^D_{11}|\times|\mu^D_{12}| = |\mu^D_{21}|\times|\mu^D_{22}| = |\mu^D_{31}|\times|\mu^D_{32}|, \nonumber \\
& |\mu^D_{11}|\times|\mu^D_{13}| = |\mu^D_{21}|\times|\mu^D_{23}| = |\mu^D_{31}|\times|\mu^D_{33}|,  \nonumber  \\
& |\mu^D_{12}|\times|\mu^D_{13}| = |\mu^D_{22}|\times|\mu^D_{23}| = |\mu^D_{32}|\times|\mu^D_{33}|.  \nonumber 
\end{eqnarray}
\item $|\mu^D_{ij}|\times|\mu^D_{i'k}|$, $i\neq i'$, $j\neq k$. The distinguishable $(j,k)$ pairs are $\{1,2\}, \{1,3\}, \{2,3\}$: 
\begin{eqnarray}
& |\mu^D_{11}|\times|\mu^D_{22}| =  |\mu^D_{11}|\times|\mu^D_{32}|  =  |\mu^D_{21}|\times|\mu^D_{12}|  =  |\mu^D_{21}|\times|\mu^D_{32}| =  |\mu^D_{31}|\times|\mu^D_{12}| = |\mu^D_{31}|\times|\mu^D_{22}|, \nonumber \\
& |\mu^D_{11}|\times|\mu^D_{23}| =  |\mu^D_{11}|\times|\mu^D_{33}|  =  |\mu^D_{21}|\times|\mu^D_{13}|  =  |\mu^D_{21}|\times|\mu^D_{33}| =  |\mu^D_{31}|\times|\mu^D_{13}| = |\mu^D_{31}|\times|\mu^D_{23}|, \nonumber \\
& |\mu^D_{12}|\times|\mu^D_{23}| =  |\mu^D_{12}|\times|\mu^D_{33}|  =  |\mu^D_{22}|\times|\mu^D_{13}|  =  |\mu^D_{22}|\times|\mu^D_{33}| =  |\mu^D_{32}|\times|\mu^D_{13}| = |\mu^D_{32}|\times|\mu^D_{23}|. \nonumber
\end{eqnarray}
\end{itemize}

\begin{figure}[t]
\begin{center}
\includegraphics[angle=0,width=1\textwidth]{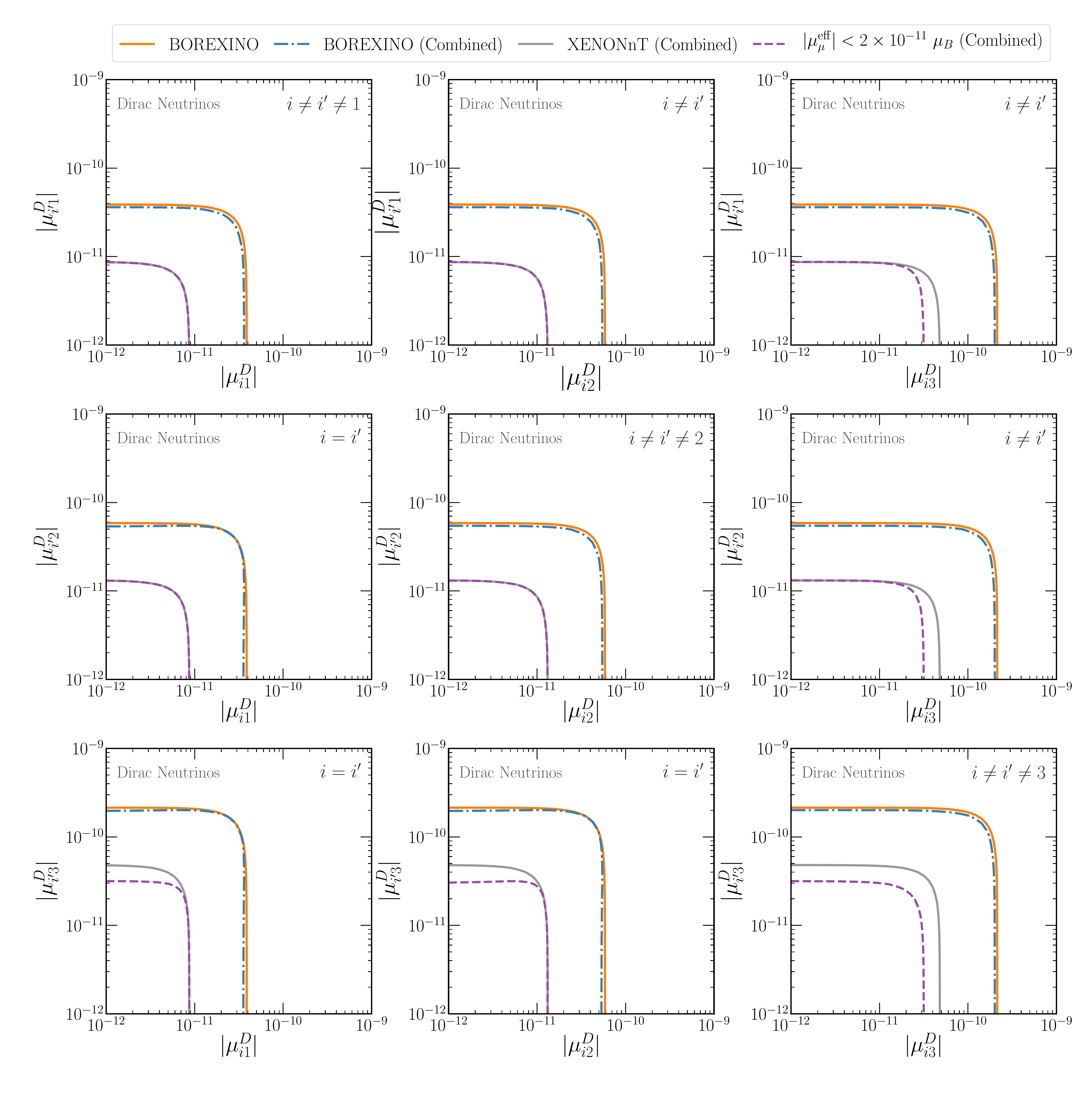}
    \vspace*{-1.2cm}
    \caption{Dirac neutrinos. 90\% C.L. allowed regions in all
    $|\mu^{D}_{ij}|\times|\mu^{D}_{i'k}|$-planes, extracted from different subsets of existing and hypothetical future data. `BOREXINO (Combined)' stands for data from Borexino, GEMMA, and LSND. `XENONnT (Combined)' stands for data from XENONnT, Borexino, GEMMA, and LSND. `$|\mu^{\rm eff}_{\mu}|<2\times 10^{-11}~\mu_B$ (Combined)' stands for data from a future experiment that constraints $|\mu^{\rm eff}_{\mu}|<2\times 10^{-11}~\mu_B$. See Section~\ref{sec:experiments} for details. In all panels, $i,i'=1,2,3$, along with the constraint in the top right-hand corner.}   
    \label{fig:dirac_magplots_all}
\end{center}
\end{figure}

Fig.~\ref{fig:dirac_magplots_all} depicts the constraints on all distinguishable (in principle) pairs of $|\mu^D_{ij}|,|\mu^D_{i'k}|$, in the corresponding $|\mu^D_{ij}|\times |\mu^D_{i'k}|$-planes. The different curves correspond to the constraints imposed by Borexino (orange contour), Borexino data combined with LSND and GEMMA (blue, dot-dashed contour), and XENONnT combined with Borexino, LSND, and GEMMA (grey contour). The dashed, purple line, corresponds to a hypothetical future bound, obtained by combining the existing XENONnT data with a future experiment that constrains $|\mu^{\rm eff}_{\mu}|<2\times 10^{-11}~\mu_B$ at the 90\% C.L.
%%%
%%%

When it comes to existing constraints on $|\mu^D_{ij}|,|\mu^D_{i'k}|$ pairs, as expected, the constraints from solar data also overwhelm those of all Earth-bound experiments, especially once one considers the very recent results reported by XENONnT.  The situation is different once one includes future constraints from an experiment sensitive to $|\mu^{\rm eff}_{\mu}|^2$. The impact of these, already discussed in the context of upper bounds on individual $|\mu^D_{ij}|$ around Table~\ref{tab:DirBnds}, can be clearly seen in Fig.~\ref{fig:dirac_magplots_all}, in the planes that involve the $|\mu^D_{i3}|$ elements.

In the far future, assuming experiments are restricted to measuring $|\mu^{\rm eff}_{\alpha}|^2$ and $|\mu^{\rm eff}_{i}|^2$ (and even different versions of the ``oscillated'' $|\mu^{\rm eff}_{\alpha}|^2$), data will still only depend on the $\mu_{ij}^D$ through $S_{jk}$. This means that there are several $\mu_{ij}^D$ ``subsets'' that are indistinguishable from one another and from the most general case. To explore this further, we define the complex 3-component vector $\vec{v}_j=(\mu^D_{1j},\mu^D_{2j},\mu^D_{3j})$, $j=1,2,3$, so $S_{jk}=\vec{v}_j\cdot\vec{v}_k^*$. All observables are proportional to the dot-products of the three different vectors $\vec{v}$ and hence do not depend on rigid rotations in the (complex) space defined by the $\vec{v}_j$. This rotational symmetry is the one we had been exploring above. Taking advantage of this invariance, we can, for example, choose the 1-direction such that $\vec{v}_1=(\mu^{D\star}_{11},0,0)$ and the 2-direction such that $\vec{v}_2=(\mu^{D\star}_{12},\mu^{D\star}_{22},0)$.\footnote{The $\star$ is mean to indicate that these are not entries of a generic matrix but one where some of the elements are known to vanish.} There is no freedom to reduced the number of components of the third vector, $\vec{v}_3=(\mu^{D\star}_{13},\mu^{D\star}_{23},\mu^{D\star}_{33})$. The entire $\mu^{D}_{ij}$ parameter space -- 9 complex parameters -- can be perfectly mimicked by a reduced parameter space -- 6 complex parameters -- where $\mu^D_{21},\mu^D_{31},\mu^D_{32}$ vanish exactly. Hence, several (as many as we can imagine) idealized measurements of $|\mu^{\rm eff}_{\alpha}|^2$ and $|\mu^{\rm eff}_{i}|^2$ may well be able to establish that neutrinos have a magnetic moment, but they cannot reveal whether, for example, some of them vanish.

%%%%%%%%%%%%%%%%%%%%%%%%%%%%%%%%%%%%%%%%%%%%%%%%%%%%%%%%%%%%%%%%%%%%%%%%%%

%%%%%%%%%%%%%%%%%%
\section{Conclusions}
\label{sec:conclusions}
\setcounter{equation}{0}
%%%%%%%%%%%%%%%%%%

Massive neutrinos are guaranteed to have nonzero electromagnetic moments. The sizes of these dipole moments are functions of all neutrino interactions with known and unknown particles and depend on the nature of the neutrino -- Majorana fermion versus Dirac fermion. 

Since there are at least three neutrino families, the neutrino dipole moments define a matrix. The number of independent electromagnetic moments depends on the number of neutrino families and the nature of the neutrinos. Here, we estimated the current upper bounds on all independent neutrino electromagnetic moments, concentrating on Earth-bound experiments and measurements with solar neutrinos. We considered the hypotheses that neutrinos are Majorana fermions or Dirac fermions. Our results, obtained after marginalizing over all other dipole-moment observables (magnitudes and phases), are listed in  Tables~\ref{tab:MajBnds} and \ref{tab:DirBnds}. 
We included the very recent results reported by the XENONnT experiment, sensitive to $pp$-solar neutrinos. 
Right now, XENONnT data provide the most stringent bounds on all elements of the neutrino electromagnetic moment matrix, independent from the nature of the neutrinos. 
This was already true of published solar neutrino data from the Borexino experiment, which makes use of the scattering of $^7$Be solar neutrinos.  

For the same number of neutrino families, there are more independent neutrino electromagnetic dipole moments if neutrinos are Dirac fermions. This translates into weaker bounds on the magnitudes of the elements of the dipole moment matrix relative to those obtained if neutrinos are Majorana fermions. 
As a concrete example, for Dirac neutrinos, if $|U_{e3}|^2$ were zero, solar data would be unable to constrain the magnitudes of three of the nine independent elements of the electromagnetic moment matrix. 
The situation is very different for Majorana neutrinos. 
In this case, the dependence on $|U_{e3}|^2$ of existing solar bounds is almost negligible. 

Another consequence of the Majorana fermion versus Dirac fermion distinction is that the potential physics reach of next-generation experiments depends on the nature of the neutrino. 
Here, we concentrated on a next-generation experiment that is sensitive to the neutrino electromagnetic moments via $\nu_{\mu}$ elastic scattering. 
An experiment sensitive to $|\mu^{\rm eff}_{\mu}|>2\times 10^{-11}\mu_B$ may discover that the neutrino electromagnetic moments are nonzero \emph{if neutrinos are Dirac fermions}. 
Instead, if neutrinos are Majorana fermions, such a discovery is ruled out by existing solar neutrino data, \emph{unless there are more than three light neutrinos}.  

The Majorana fermion versus Dirac fermion distinction can be effectively erased if there are more than three light neutrinos. 
For example, five Majorana neutrinos (e.g., three mostly active and two mostly sterile) allow for ten complex electromagnetic dipole moments, a good match (with one dipole moment to spare) to the nine complex electromagnetic dipole moments required to describe the couplings of three Dirac neutrinos. 
It is not clear whether these two scenarios can be disentangled, even if one assumes a large collection of very precise future experiments, including measurements of $|\mu^{\rm eff}_{e,\mu,\tau}|^2$ from the elastic scattering of all three flavor eigenstates along with different linear combinations of $|\mu^{\rm eff}_{1,2,3,\ldots}|^2$ from the scattering of solar neutrinos of different energies. %Even then, it is not at all clear one can distinguish the Majorana fermion from the Dirac fermion hypothesis for the neutrinos.

We explored in great detail what information can be acquired, in principle, on the neutrino electromagnetic moments if neutrinos are Dirac fermions. 
Unlike the Majorana case, in the Dirac case the parameter space is very large -- 9 complex parameters. 
Nonetheless, if all future information comes from measurements of $|\mu^{\rm eff}_{1,2,3,\ldots}|^2$ and $|\mu^{\rm eff}_{e,\mu,\tau}|^2$, the amount of information one can extract is much more limited than naively anticipated. 
For example, in the absence of a discovery, for a fixed value of $j=1,2,$ or 3, upper limits on $|\mu^D_{ij}|$ are \emph{identical} for all $i=1,2,3$. 
Similarly, excluded regions in several $\mu_{ij}\times \mu_{i'k}$ planes are also identical, and the argument persists for ``higher-dimensional'' allowed regions in $\mu_{ij}\times \ldots \times \mu_{i'k}$ spaces. 
In the case of the reduced two-dimensional $\mu_{ij}\times \mu_{i'k}$ spaces, we showed there are only nine independent excluded regions. All other 27 are related to those nine. 

The situation would be qualitatively different if the scattered neutrinos from the detection process were also, somehow, measured. 
This requires experimental capabilities that are way out of current reach. 
For example, one may consider the dipole-moment mediated process $\nu_{\alpha}+e^-\to\nu_{\beta}+e^-$, $\alpha,\beta=e,\mu,\tau$. Assuming a left-handed-helicity $\nu_{\alpha}$ and neutrino energies much larger than the neutrino masses -- guaranteed of all available neutrino beams --  the outgoing $\nu_{\beta}$ would have right-handed helicity. 
If neutrinos are Dirac fermions, the observation of the right-handed-helicity $\nu_{\beta}$ requires chirality violation and is hence very efficiently suppressed by the neutrino masses squared (in units of the neutrino energy). For all practical purposes, right-handed-helicity  $\nu_{\beta}$ are sterile neutrinos. Instead, if neutrinos are Majorana fermions, the right-handed-helicity $\nu_{\beta}$ would behave as what is casually referred to as a $\bar{\nu}_{\beta}$ and, if measured via charged-current weak interactions, would lead to the production of an $\ell_{\beta}^+$. In the latter scenario, not only would one be able to measure $\mu_{\alpha\beta}$ (as opposed to $\mu^{\rm eff}_{\alpha}$), but one would also have discovered that lepton-number-symmetry is violated and that neutrinos are Majorana fermions. 

The fact that experimental constraints on the neutrino electromagnetic moments are weaker (and the discovery potential, in some sense, stronger) if  neutrinos are Dirac fermions is orthogonal to theoretical expectations that point to a strong correlation between potentially large neutrino electromagnetic moments and Majorana fermions \cite{Bell:2005kz,Bell:2006wi}, highlighted in the Introduction. 
The discovery of neutrino electromagnetic moments of order $10^{-11}\mu_B$, coupled to knowledge that neutrinos are Dirac fermions, would indicate that the robust assumptions made in \cite{Bell:2005kz,Bell:2006wi} do not apply and that the physics behind nonzero neutrino masses is more puzzling and subtle than the community currently suspects.

%%%%%%%%%%%%%
\section*{Acknowledgements}
This research was supported in part through the computational resources and staff contributions provided for the Quest high performance computing facility at Northwestern University which is jointly supported by the Office of the Provost, the Office for Research, and Northwestern University Information Technology.
It was also supported in part by the US Department of Energy (DOE) grant \#DE-SC0010143 and in part by the National Science Foundation under Grant No.~PHY-1630782. The work of ZT is supported by the Neutrino Theory Network Program Grant \#DE-AC02-07CHI11359 and the US DOE under award \#DE-SC0020250.
AdG and ZT are thankful to the KITP in Santa Barbara, where part of this work was pursued, for its hospitality and productive atmosphere. The research at KITP was supported in part by the National Science Foundation under Grant No. NSF PHY-1748958.
The document was prepared using the resources of the Fermi National Accelerator Laboratory (Fermilab), a DOE, Office of Science, HEP User Facility. Fermilab is managed by Fermi Research Alliance, LLC (FRA), acting under Contract No. DE-AC02-07CH11359.
%%%%%%%%%%%%%

\bibliographystyle{utphys}
\bibliography{References}

\end{document}